\documentclass[aps,pre,twocolumn,groupedaddress,superscriptaddress,showpacs]{revtex4-1}
\usepackage{amsmath}
\usepackage{graphicx}
\usepackage{float}
\usepackage{xcolor}

\begin{document}

  \title {Crossover from three- to six-fold symmetry of colloidal aggregates in circular traps}

  \author{T. Geigenfeind}
   \email{thomas.geigenfeind@uni-bayreuth.de}
    \affiliation{Theoretische Physik II, Physikalisches Institut, Universit\"{a}t Bayreuth, D-95440 Bayreuth, Germany}
    
  \author{C. S. Dias}
   \email{csdias@fc.ul.pt}
    \affiliation{Departamento de F\'{\i}sica, Faculdade de Ci\^{e}ncias, Universidade de Lisboa, 
    1749-016 Lisboa, Portugal}
    \affiliation{Centro de F\'{i}sica Te\'{o}rica e Computacional, Universidade de Lisboa, 
    1749-016 Lisboa, Portugal}
    
  \author{M. M. Telo da Gama}
   \email{mmgama@fc.ul.pt}
    \affiliation{Departamento de F\'{\i}sica, Faculdade de Ci\^{e}ncias, Universidade de Lisboa, 
    1749-016 Lisboa, Portugal}
    \affiliation{Centro de F\'{i}sica Te\'{o}rica e Computacional, Universidade de Lisboa, 
    1749-016 Lisboa, Portugal}

  \author{D. de las Heras}
   \email{delasheras.daniel@gmail.com}
    \affiliation{Theoretische Physik II, Physikalisches Institut, Universit\"{a}t Bayreuth, D-95440 Bayreuth, Germany}
      
  \author{N. A. M. Ara\'ujo}
   \email{nmaraujo@fc.ul.pt}
    \affiliation{Departamento de F\'{\i}sica, Faculdade de Ci\^{e}ncias, Universidade de Lisboa, 
    1749-016 Lisboa, Portugal}
    \affiliation{Centro de F\'{i}sica Te\'{o}rica e Computacional, Universidade de Lisboa, 
    1749-016 Lisboa, Portugal}

\begin{abstract}
At sufficiently low temperatures and high densities, repulsive spherical
	particles in two-dimensions (2d) form close-packed structures with six-fold symmetry. By
	contrast, when the interparticle interaction has an attractive anisotropic component, the 
	structure may exhibit the symmetry of the interaction. We consider a suspension of 
	spherical particles interacting through an isotropic repulsive potential and a three-fold
	symmetric attractive interaction, confined in circular potential traps in 2d. We
	find that, due to the competition between the interparticle and the external potentials, 
	the particles self-organize into structures with three- or six-fold symmetry, depending on 
	the width of the traps. For intermediate trap widths, a core-shell structure is formed, where
	the core has six-fold symmetry and the shell is three-fold symmetric. When the width
	of the trap changes periodically in time, the symmetry of the colloidal structure also 
	changes, but it does not necessarily follow that of the corresponding static trap.
\end{abstract}

  \maketitle

\section{Introduction}

The control over the self-organization of colloidal particles is a problem of both fundamental and practical interest 
\cite{Sacanna2011,Chen2011,Romano2011a,Nykypanchuk2008,Parak2011,Yang2008,Furst2011,Manoharan2015,Duguet2011,Dias2017}.
One promising approach is the use of chemically or otherwise patterned substrates, where the particle-substrate interaction is spatially dependent and may even change with time.
Spatial patterns on substrates may be created using, for example, lithographic
methods~\cite{Ramsteiner2009,Wang2004a}, 
magnetic domains~\cite{Tierno2014,Loehr2017}, chemical coating \cite{Aizenberg1999,Chen2000,Guo2001}, or DNA-mediated functionalization of interfaces \cite{Joshi2016}.

It has been shown that the equilibrium phases of colloidal particles are affected strongly by substrates 
with spatial patterns. For example, patterns may induce new surface phases~\cite{Heni2000} and crystalline structures \cite{Harreis2002},
or affect the wetting properties of the surfaces \cite{Bauer1999a}. The collective (non-equilibrium) dynamical properties are also affected. 
For example, in the limit of irreversible adsorption, a simple pattern of pits distributed in a 
square-lattice arrangement induces either local or long-range order, depending
on the size of the pits and the distance between them~\cite{Cadilhe2007}. 

For simplicity, most of the previous works have considered isotropic particles but, in general, the interparticle interaction is anisotropic. 
Anisotropy may result, for example, from the individual particle shape \cite{Sacanna2013a,Wolters2015,Glotzer2012,Yunker2013,Dias2018a}, 
heterogeneous distribution of charges \cite{Ilg2011,Nych2013,Klapp2016}, 
or functionalization of the particle surface \cite{Wang2012,Smallenburg2013a,Bianchi2011a,Dias2014a,Sokolowski2014,Syk2016,Yi2013,He2012,Kraft2011,Cates2013,Geigenfeind2016}. 
In these cases, the final structures should depend 
not only on the symmetries of the pattern but also on those of the interparticle potential. In a recent study the equilibrium properties of
particles with three-fold symmetric attractive interaction adsorbed on patterned substrates were considered~\cite{Treffensta2018}. The properties of the
pattern strongly affect both the percolation properties and the type of network in which the particles self-assemble.

Here, we investigate how the dynamics of self-organization of colloidal particles with anisotropic interparticle interactions is affected by the presence of spatial patterns. 
These patterns result from a square lattice arrangement of (Gaussian) attractive traps with a characteristic width on an otherwise flat substrate. 
We show that, the structure of the colloidal aggregates, on the substrate, depends strongly on the width of the traps. 
We consider also traps with a time dependent width and show that the dynamics may differ significantly from that of the static traps.

We introduce the model and the relevant parameters in Sec.~\ref{sec.model}. In Sec.~\ref{sec.results} we present
the results and we draw some conclusions in Sec.~\ref{sec.conclusions}.

\section{Model}\label{sec.model}

\begin{figure*}[t]
\begin{center}
\includegraphics[width=1.8\columnwidth]{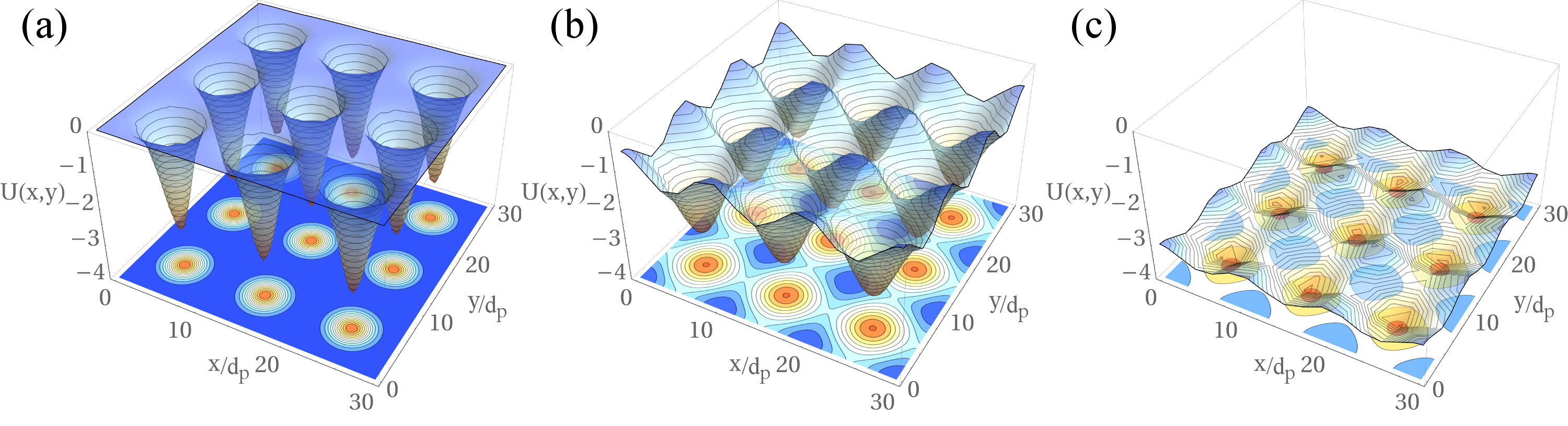}\\
\caption{Schematic representation of the attractive Gaussian potential traps distributed in a square-lattice arrangement. 
The traps are separated by a distance of ten particle diameters ($d_p$) and 
have a width (a) $R_\mathrm{W}=2$, (b) $R_\mathrm{W}=4$, and (c)
$R_\mathrm{W}=6$ in units of the particle diameter ($d_p$). The potential strength is in units of $\epsilon$.\label{fig.potentials}}
\end{center}
\end{figure*}

We consider a monodisperse suspension of spherical (colloidal) particles, where the particle-particle interaction is a superposition of an isotropic 
repulsion and a three-fold symmetric attraction. Following Ref.~\cite{Dias2016}, we describe this pairwise interaction by 
decorating the surface of the spherical particles with three patches
distributed along the equator. 
The patch-patch attractive interaction has a Gaussian form given by, 
\begin{equation}
U_{\text{patch/patch}}(r_p)=-\epsilon\exp\left[-(r_p/\sigma)^2\right],
\label{eq.gaussian} 
\end{equation}
where $r_p$ is the distance between the center of the patches, $\epsilon$ is
the interaction strength that sets the energy scale, and $\sigma=0.1$ the width of the Gaussian in units of the
effective particle diameter $d_p$ (which sets the
length scale).

The core-core interaction is repulsive and given by, 
\begin{equation}
U_{\text{part/part}}(r)=\frac{A}{k}\exp{\left[-k\left(r-d_p\right)\right]}, 
\label{eq.yukawa}
\end{equation}
where $r$ is the distance between the center of the particles, 
$A=0.25$ (in units of $\epsilon/d_p$), and $k=0.4$ is the screening length (in units of $d_p$).

To confine the particles to the surface of a planar substrate, we implemented
the method described in Ref.~\cite{Dias2016}. The surface pattern consists of 
attractive potential traps, distributed in a square-lattice
arrangement (see Fig.~\ref{fig.potentials}), with a Gaussian form,
\begin{equation}
	U_{\text{part/trap}}(r)=-3\epsilon\exp\left[-(r/R_\mathrm{W})^2\right],
	\label{eq.gaussian_well} \end{equation}
where $\epsilon$ is the strength of the patch-patch interaction (see
Eq.~(\ref{eq.gaussian})) and $R_\mathrm{W}$ is the width (range) of the trap.
The potential is truncated at a distance of $10$ particle diameters ($d_p$) from the
center of the trap.  As shown in Fig.~\ref{fig.potentials}, although the
minimum of the traps is kept fixed, the effective potential landscape
depends not only on $R_W$, but also on the distance between the center of the
traps, as in some regions the particles interact simultaneously with more
than one trap.  Since the particle-trap interaction is always attractive, this
implies that the net force acting on a particle is lower if the traps overlap. We
impose periodic boundary conditions along the x- and y-directions.

To resolve the trajectory of individual particles, we perform Langevin dynamics using the Large-scale Atomic/Molecular Massively 
Parallel Simulator (LAMMPS) \cite{Plimpton1995}. Particles are spherical with mass $m$ and inertia $I$ and the patches on their surface have negligible mass. The  
translational and rotational motion of the particles is described by the following equations, 
\begin{equation}
 m\dot{\vec{v_i}}(t)=-\nabla_{\vec{r_i}} U-\frac{m}{\tau_t}\vec{v_i}(t)+\sqrt{\frac{2mk_BT}{\tau_t}}\vec{\xi_t^i}(t), \label{eq.Langevin_dynamics_trans} 
\end{equation}
and
\begin{equation}
 I\dot{\vec{\omega_i}}(t)=-\nabla_{\vec{\theta_i}} U-\frac{I}{\tau_r}\vec{\omega_i}(t)+\sqrt{\frac{2Ik_BT}{\tau_r}}\vec{\xi_r^i}(t), \label{eq.Langevin_dynamics_rot}
\end{equation}
where, $\vec{v_i}$ and $\vec{\omega_i}$ are the translational and angular velocities of particle $i$. The translational and rotational damping times are 
given by $\tau_t=0.02\sqrt{(m d_p^2/\epsilon)}$ and $\tau_r=10\tau_t/3$ for spherical colloids. 
$\vec{\xi_t^i}(t)$ and $\vec{\xi_r^i}(t)$ are stochastic terms taken from a truncated random distribution of zero mean
and standard deviation of one unit~\cite{Dunweg1991}. 
 $U$ is the total potential with contributions from the particle-particle and
particle-trap interactions, and therefore it depends on both the positions $\vec{r_i}$ and orientations
$\vec{\theta_i}$ of all particles $i=1...N$ in the system. Note that, although the particles are on a planar substrate, they
can still rotate in three dimensions.

\begin{figure}[t]
\begin{center}
\includegraphics[width=1\columnwidth]{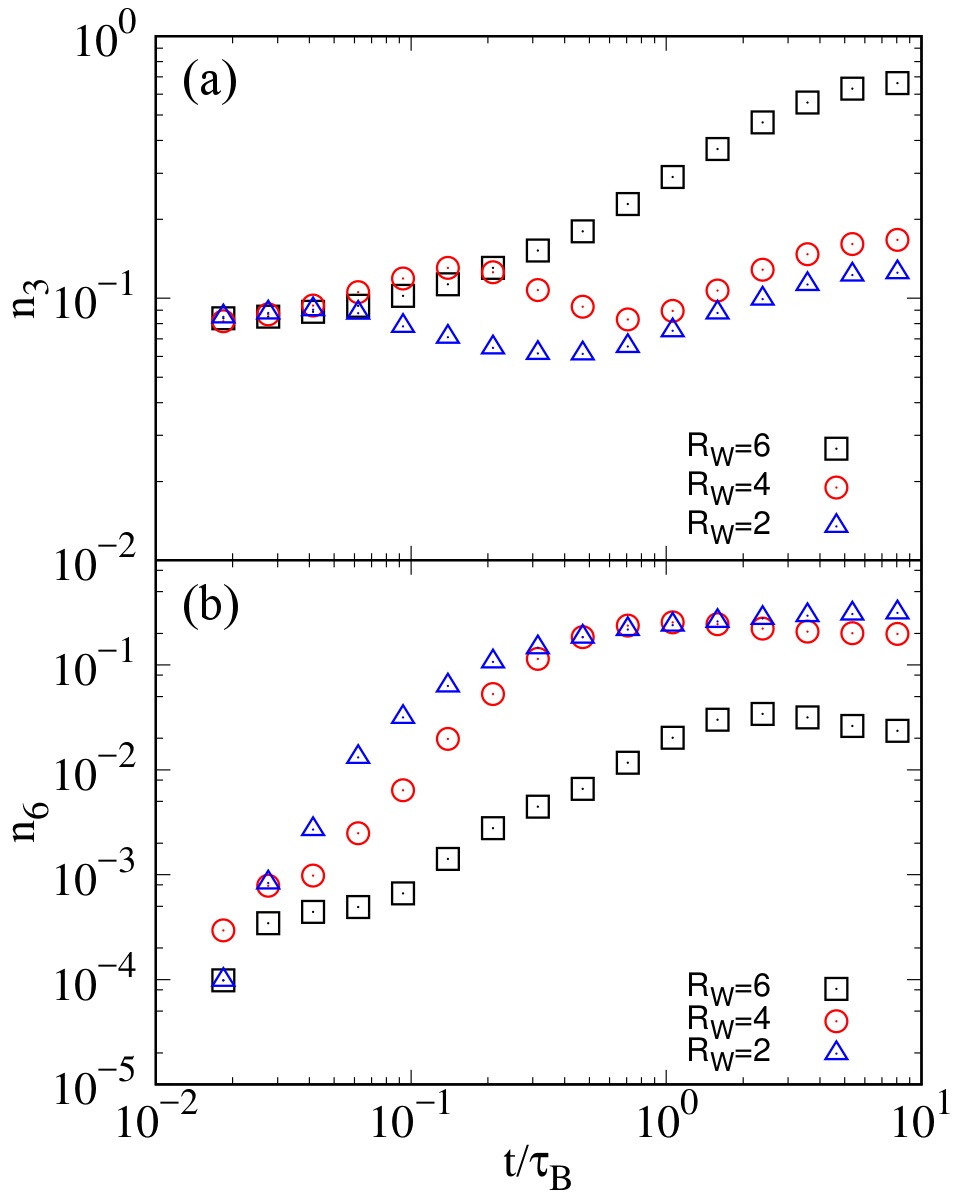}\\
\caption{Time dependence of the fraction of particles with a local (a) three- ($n_3$)
	and (b) six-fold ($n_6$) bond order parameter above $1/2$ for
	different widths of the traps, namely, 
$R_\mathrm{W}=\{2,4,6\}$. Simulations are performed on a substrate of linear size
	$L=40$ (in units of particle diameter $d_p$). \label{fig.phi_dynamics}}
\end{center}
\end{figure}

\section{Results}\label{sec.results}

Particles are initially distributed, without overlapping, uniformly at random
on the substrate with a given particle number density $\rho$, defined as the
number of particles per unit volume.  Simulations were performed at a reduced
temperature $T^*=k_BT/\epsilon$, where $T$ is the thermostat temperature,
$\epsilon$ the strength of the patch-patch interaction and $k_B$ the Boltzmann
constant.  Unless otherwise stated, we rescale the time by the Brownian time
$\tau_\mathrm{B}=d_p^2/D_t$, where $D_t$ is the translational diffusion
coefficient $D_t=k_BT\tau_t/m$. $\tau_\mathrm{B}$ is related
to the typical time taken by a particle to diffuse in an area $d_p^2$,
considering an overdamped regime \cite{Dias2018}.
All results are averages over (i) 10 different
realizations and (ii) all traps in each realization.

To evaluate the local structure formed by the colloidal particles, we measure
the local $k$-fold bond order parameter of the $i$-th particle~\cite{Nunes2018},
\begin{equation}
  \phi^{(i)}_{k=\{3,6\}}=\left|\frac{1}{\max\{N_l,k\}}\sum^{N_l}_{j=1}e^{-ik\theta_{ij}}\right|,
\end{equation}
where $N_l$ is the number of neighbors around the particle within a cut-off
radius $r_{\text{cut}}=1.3$ in units of the particle diameter ($d_p$). $\theta_{ij}$ is the angle
between the vector connecting particles $i$ and $j$ and the x-direction (parallel to
the substrate). $k$ is a parameter related to the local symmetry such that, $\phi_6$
is one for perfect six-fold symmetry and $\phi_3$ is one for perfect three-fold symmetry.  We
then define $n_k$ as the fraction of particles with a value of $\phi_k$ above
a specific threshold that we set to $1/2$.

In what follows, we analyze first the dynamics at a temperature $T^*=0.0625$ which is
below the (3D) gas-liquid coexistence curve.  Then, we consider traps with a
time-dependent width at temperatures
$T^*=\{0.0625,0.075,0.1,0.125\}$, corresponding to temperatures below and
above the bulk gas-liquid coexistence curve, as predicted by the phase diagram
of Ref.~\cite{Dias2018b}. 

\begin{figure}[t]
\begin{center}
	\includegraphics[width=1\columnwidth]{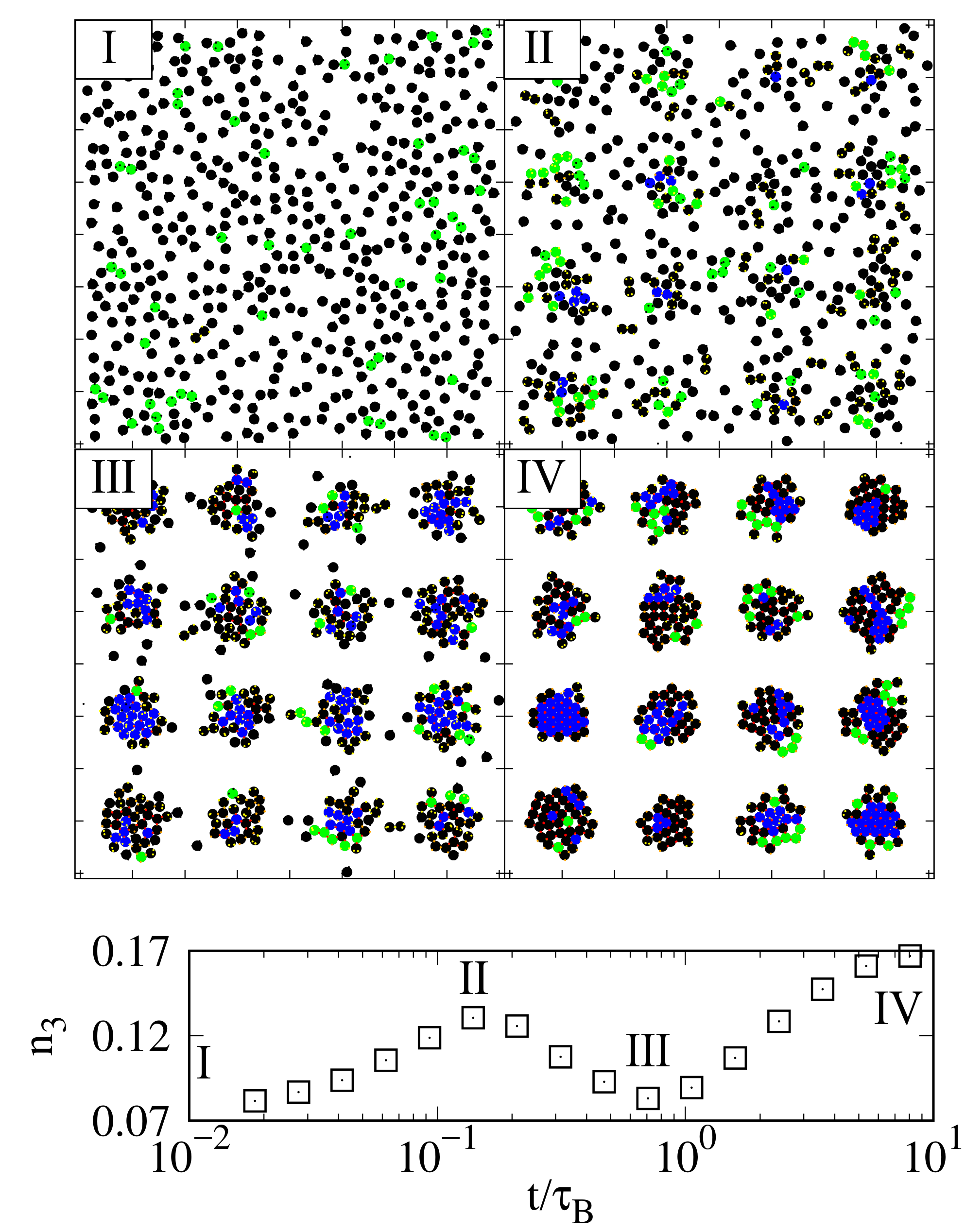}\\
\caption{
        Bottom: Time dependence of the fraction of particles ($n_3$) with a
	local three-fold bond order parameter above $1/2$ for traps with
	$R_\mathrm{W}=4$ and particle number density $\rho=0.25$. Simulations
	are performed on a substrate of linear size $L=40$. 
        Top: Snapshots of the structure at
	different times (as marked in the bottom plot). Blue and green
	particles have a local six- and three-fold bond order parameter above $1/2$,
	respectively.  In black are the particles with none of the two local
	bond order parameters above the threshold. \label{fig.phi3_dynamics}
        }
\end{center}
\end{figure}

\subsection{Constant potential traps}
Figure~\ref{fig.phi_dynamics} shows the time evolution of the fraction of
particles with a local (a) three- ($n_3$) and (b) six-fold ($n_6$) symmetry,
for different values of the width of the traps ($R_\mathrm{W}=2,4,6$) but the same
initial particle number density ($\rho=0.25$). Let us focus on the case
$R_\mathrm{W}=6$ (squares), which is the largest value that we have considered.
Both $n_3$ and $n_6$ increase in time as the particles accumulate in the
potential traps.  However, the asymptotic value of $n_3$ is about two orders of
magnitude larger than $n_6$, as most particles have a local three-fold
symmetry, in line with the symmetry of the patch-patch attractive potential. By contrast,
when the width of the trap is reduced, the values of $n_3$ and $n_6$ are
comparable. For $R_\mathrm{W}=2$, the fraction of particles with a local
six-fold symmetry is even larger than that of particles with a three-fold
symmetry (i.e., $n_6>n_3$). This can be explained by the following mechanism.
Due to the particle-trap interaction, particles are dragged towards the center
of the potential traps, increasing the local density there. The dragging forces are 
stronger at lower values of $R_\mathrm{W}$ (see
Eq.~(\ref{eq.gaussian_well})). Thus, while for $R_\mathrm{W}=6$ the symmetry of
the aggregates resembles that of the particle-particle attractive potential,
for $R_\mathrm{W}=2$, the attractive particle-trap forces favor an increase in
the local density (packing), and a six-fold symmetry emerges.

\begin{figure}[t]
	\begin{center}
	\includegraphics[width=1\columnwidth]{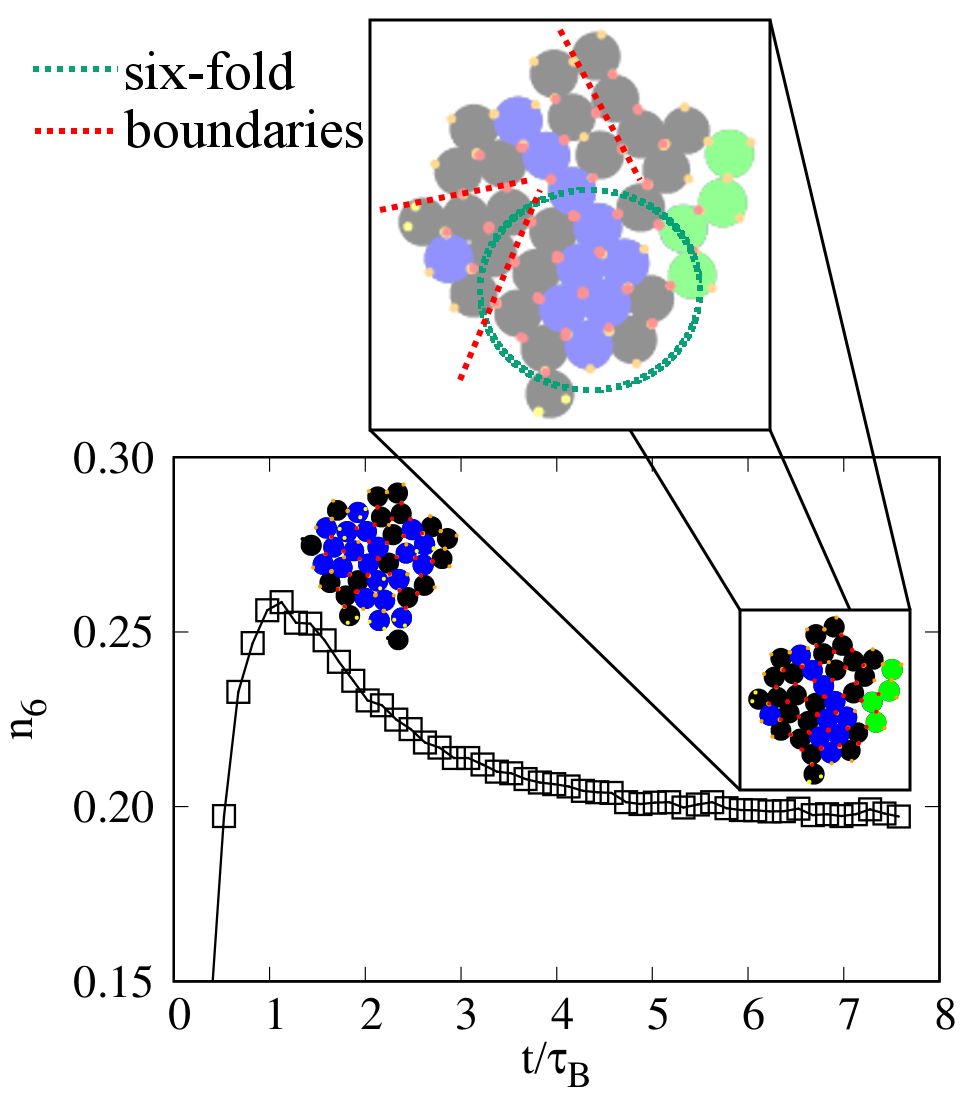}\\
\caption{Time dependence of the fraction of particles ($n_6$) with a local
		six-fold bond order parameter above $1/2$ for traps with
		$R_\mathrm{W}=4$ and particle number density $\rho=0.25$.
		Simulations are performed on a substrate of linear size $L=40$.  Snapshots are examples
		of clusters at the point where the maximum occurs (left) and
		at the end of the simulation (right). Blue and green particles
		have a local six- and three-fold bond order parameter above $1/2$,
		respectively.  In black are the particles where none of the
		two local order parameters are above threshold. The zoomed region at
		the top shows the lines of defects (red-dotted line) between
		six-fold regions (green-dotted line).
		\label{fig.phi6_dynamics}}
	\end{center} 
\end{figure}

Figure~\ref{fig.phi_dynamics} shows that the dynamics for traps with
$R_\mathrm{W}=4$ exhibits a non-monotonic behavior of $n_3$.
Figure~\ref{fig.phi3_dynamics} depicts snapshots of the structure at different
instants, as pointed out in the plot in the bottom of the same figure.
Initially (in I), the particles are randomly distributed in space, without
overlapping. As the potential traps are switched on, the particles are
dragged towards the center of the traps, establishing bonds with other
particles. The value of $n_3$ increases (from I to II), since the
particle-particle attractive interaction favors three bonds per particle (green
particles). As more particles are attracted to the traps (from II to III), the
fraction of particles with six neighbors in the center of the trap (blue
particles) increases and the value of $n_3$ decreases. As the aggregates
grow (from III to IV), the outer particles are under weaker trap forces
than the inner ones, favoring again the particle-particle bonds over packing.

\begin{figure}[t]
\begin{center}
\includegraphics[width=1\columnwidth]{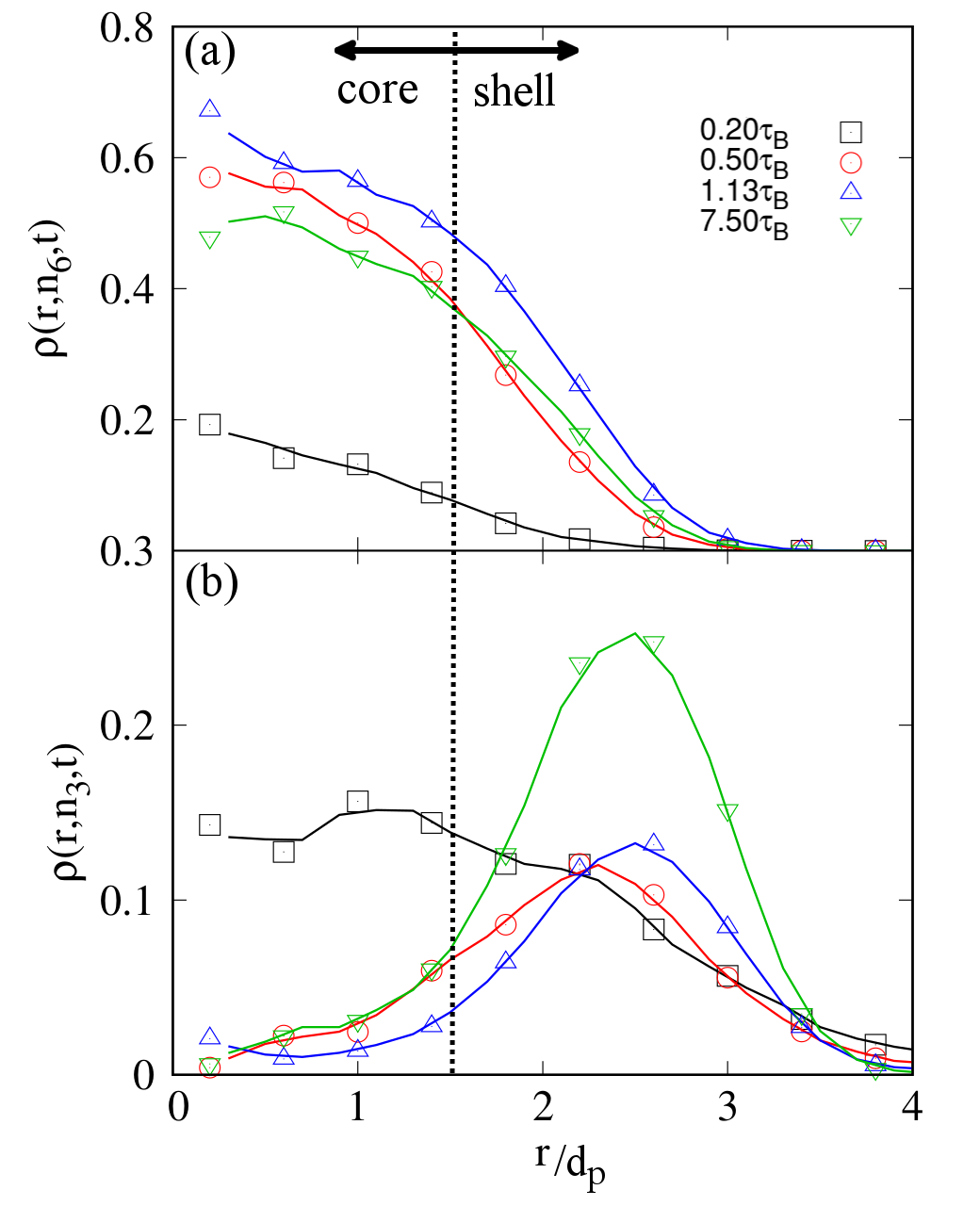}\\
\caption{(a) Number density of particles with a local six-fold bond
	order parameter above $1/2$, measured from the center of the traps at times
	$t=\{0.20, 0.50, 1.13, 7.50\}\tau_B$.  (b) Number density 
	of particles with a local three-fold bond order parameter above
	$1/2$,  measured from the center of the traps at times $t=\{0.20, 0.50,
	1.13, 7.50\}\tau_B$.  Solid lines are averages over neighboring points
	on the left and right to show the overall tendency.  Simulations were
	performed on a square substrate of size $L=40$.\label{fig.RDF}}
\end{center}
\end{figure}

When $R_\mathrm{W}=4$, the increase in $n_6$  is in fact a transient. 
As shown in Fig.~\ref{fig.phi6_dynamics}, although $n_6$ initially increases due to packing, it eventually decreases at longer times. Particles in the aggregates 
relax slowly to form domains with six-fold symmetry with strong particle-particle bonds along the grain boundaries (see inset of Fig.~\ref{fig.phi6_dynamics}). As a result, 
the fraction of particles with six-fold symmetry is reduced and the value of $\phi_k$ for a large fraction of the particles is below the threshold for both $k=3$ and $6$.

\begin{figure}[t]
\begin{center}
\includegraphics[width=1\columnwidth]{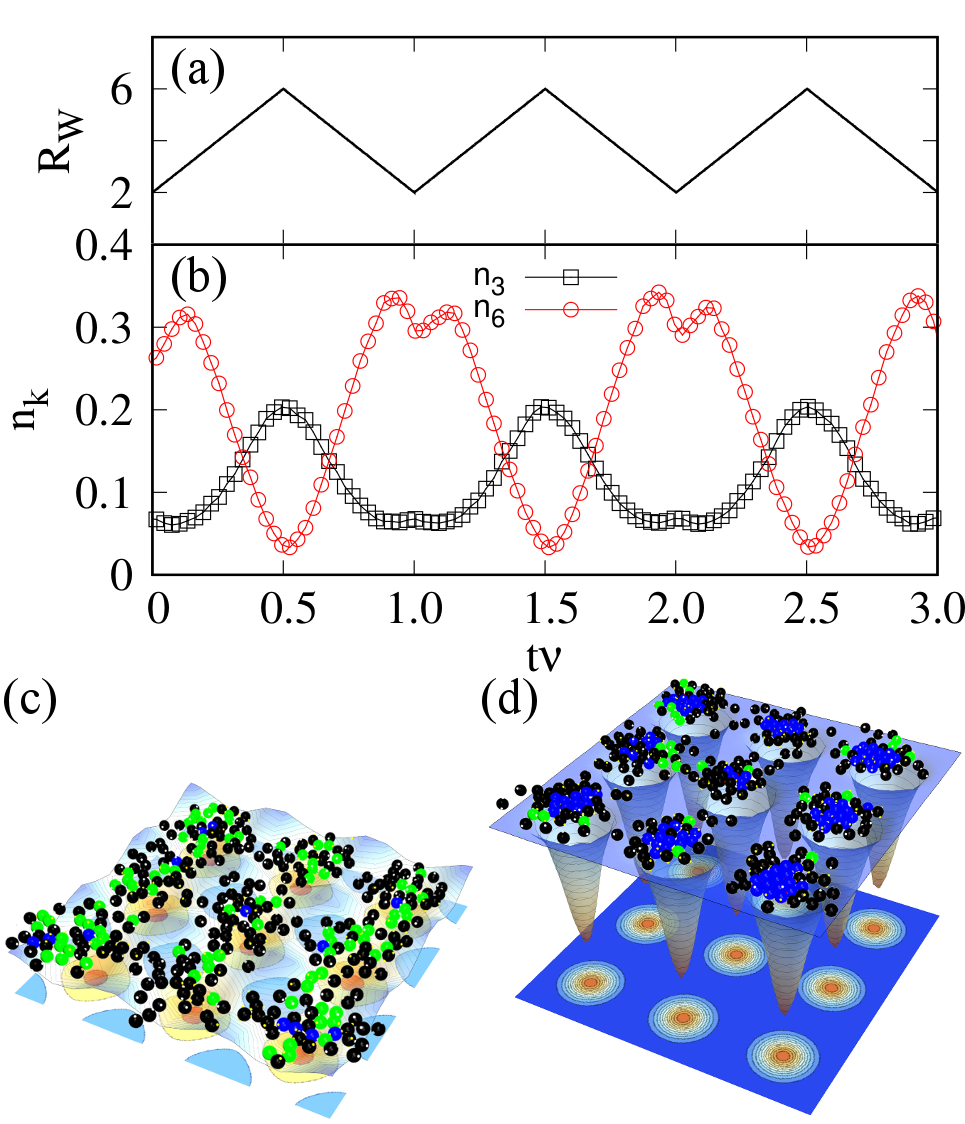}\\
\caption{(a) Time dependent width of the potential traps ($R_\mathrm{W}$) that varies periodically between two and six.
(b) Time dependence of the fraction of particles ($n_6$) with a local six-fold and ($n_3$) three-fold bond order parameter above $1/2$ for traps with the same oscillating width
$R_\mathrm{W}$, particle number density $\rho=0.3$, and reduced temperature $T^*=0.125$. Here, time is rescaled in units of the period of the oscillation.
Simulations are performed on a square substrate of size $L=40$.
Simulation snapshots at the two limiting potential widths (c) $R_\mathrm{W}=6$ and (d) $R_\mathrm{W}=2$ for the oscillating traps.
Blue and green particles have a local six- and three-fold bond order parameter above $1/2$, respectively. 
In black are the particles where none of the two local order parameters are above the threshold.
\label{fig.oscillating_well}}
\end{center}
\end{figure}


Figure~\ref{fig.RDF} shows the number density ($\rho$) of particles around
the center of the trap for $R_\mathrm{W}=4$.
Figures~\ref{fig.RDF}(a)~and~(b) show $\rho(r,n_3)$ and
$\rho(r,n_6)$, respectively. 
Note that the maxima of $n_3$ and $n_6$
occur at different positions. This difference corroborates the hypothesis that
the local three- and six-fold symmetric structures are formed in different
regions. In the center of the trap, the particles self-organize (pack) with six-fold
symmetry, while in the perimeter most particles form three bonds with other
particles.  The time evolution of the radial distribution function shows that,
at early times, both structures form near the center of the trap
independently. As time evolves, a separation of the structures is observed,
with the six-fold structure in the center of the trap and the three fold one in the 
perimeter. Asymptotically, the six-fold peak of the radial distribution
function decreases slightly due to the rearrangement discussed above.

\begin{figure}[t]
\begin{center}
\includegraphics[width=1\columnwidth]{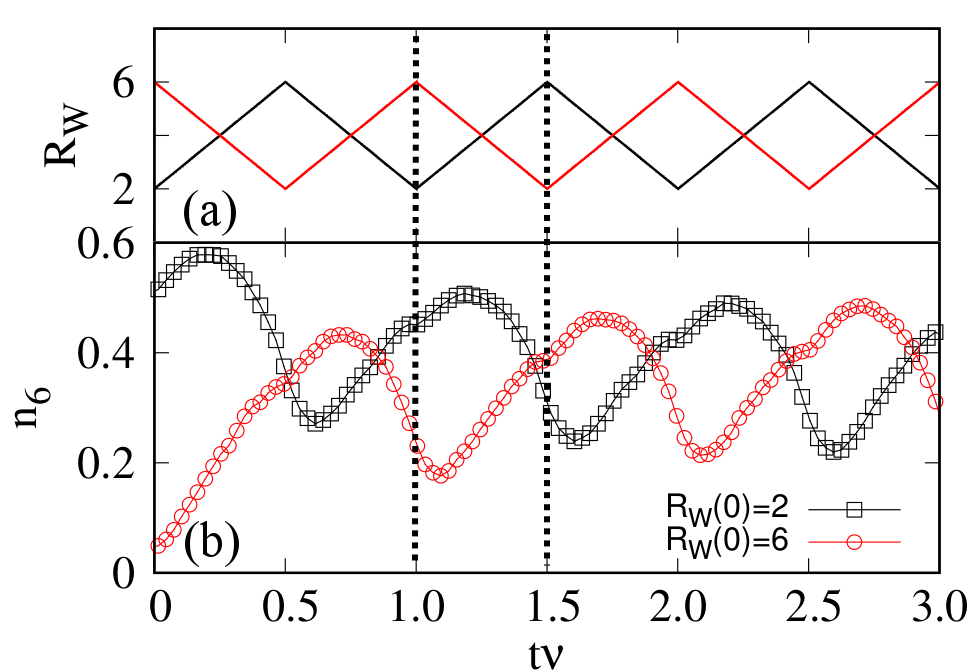}\\
 \caption{(a) Time dependent width of the potential traps
	($R_\mathrm{W}$) which varies periodically between two to six.  (b) Time
	dependence of the fraction of particles ($n_6$) with a local six-fold
	bond order parameter above $1/2$ for oscillating traps with initial widths 
	$R_\mathrm{W}(0)=2$ (squares) and $R_\mathrm{W}(0)=6$ (circles) for a particle number
	density $\rho=0.3$ and reduced temperature $T^*=0.1$.  
	Here, time is rescaled in units of the period of the oscillation.  
	Simulations are performed on a square substrate of size $L=40$. \label{fig.oscillating_wells_T4}}
 \end{center}
\end{figure}

\subsection{Time-dependent potential traps}

\begin{figure}[t]
\begin{center}
\includegraphics[width=1\columnwidth]{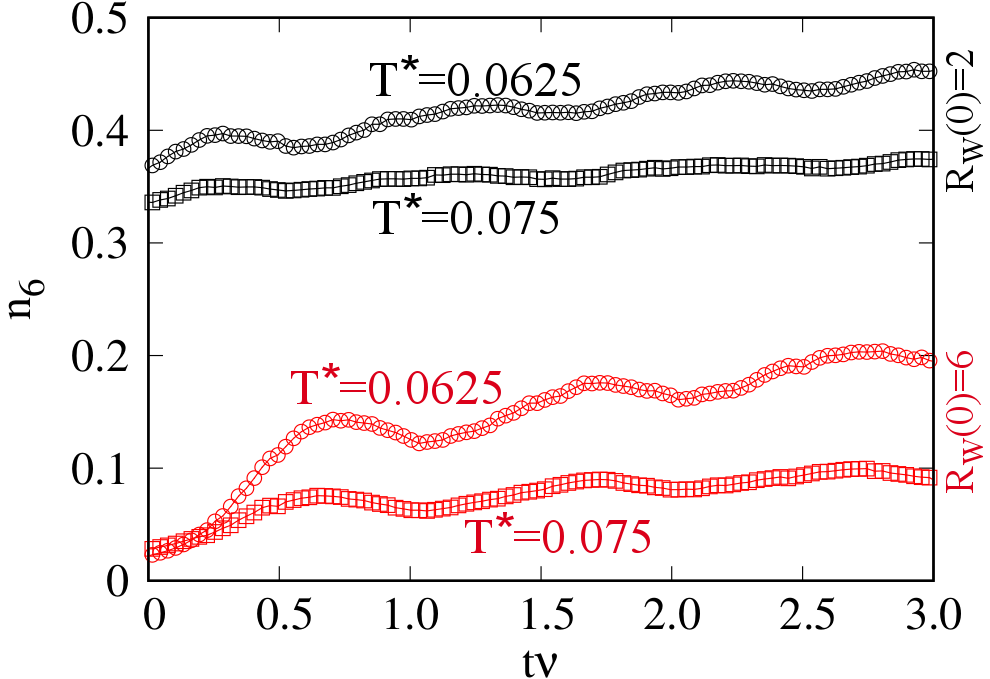}\\
 \caption{Time dependence of the fraction of particles ($n_6$) with a local six-fold bond parameter above $1/2$ for traps with $R_\mathrm{W}$ 
 varying periodically from two to six with initial trap widths $R_\mathrm{W}(0)=2$ 
 and $R_\mathrm{W}(0)=6$ (same as Fig.~\ref{fig.oscillating_wells_T4}(a)) and reduced temperatures
$T^*=0.0625$ (squares) and $T^*=0.075$ (circles), for a particle number density $\rho=0.3$. 
Here, time is rescaled in units of the period of the oscillation. 
Simulations are performed on a square substrate of size $L=40$. \label{fig.oscillating_wells_T2.5_3}}
 \end{center}
 \end{figure}
 
The results reported above suggest the design of a device, where the symmetry
of the aggregates may be dynamically switched from three to six fold through the 
width of the potential traps.  The effectiveness of such a device, however, depends 
on how the rate of the symmetry change compares to the different timescales
involved, namely those related to bond breaking/formation and
rotational/translational diffusion \cite{Dias2018,Tavares2018}.

We considered time dependent trap widths, as shown in Fig.~\ref{fig.oscillating_well}(a). 
We start with traps with
$R_\mathrm{W}=2$ and periodically increase
$R_\mathrm{W}$ linearly to $R_\mathrm{W}=6$ and then reduce it back (linearly)
to $R_\mathrm{W}=2$.  Let us discuss the dynamics at $T^*=0.125$, the
highest temperature considered here, where the oscillations have a period $\nu^{-1}=20\tau_B$. In Fig.~\ref{fig.oscillating_well}(b) we plot
the fraction of particles with a local six- and three-fold symmetry, as a
function of time. Here, for convenience, time is rescaled in units of the period of
the oscillation of $R_\mathrm{W}$ ($\nu^{-1}$).  At this temperature, the
local structure oscillates between three-fold and six-fold symmetry, with the same
frequency and phase of $R_\mathrm{W}$ (see snapshots
Figs.~\ref{fig.oscillating_well}(c)~and(d)).  A local minimum in $n_6$ is
observed at $R_\mathrm{W}=2$ indicating that the optimal
width of the trap to promote local six-fold symmetry is around three
particle diameters ($d_p$).
 
Figure~\ref{fig.oscillating_wells_T4}(b) shows the time dependence of $n_6$
(black squares) for the same oscillating $R_\mathrm{W}$ at a slightly
lower temperature ($T^*=0.1$). In this case, $n_6$ oscillates with a
period $\nu^{-1}=16\tau_B$, but the phase is shifted by approximately $1/4$ of the
period.  To investigate the dependence on the initial conditions, we applied the
same oscillating trap $R_\mathrm{W}$, starting at $R_\mathrm{W}=6$ (red
circles in Fig.~\ref{fig.oscillating_wells_T4}). The same shift is observed,
and the periodic behavior at longer times is clearly independent of the initial
conditions.

When the temperature is reduced further, namely, $T^*=\{0.0625, 0.075\}$
(corresponding to periods of $\nu^{-1}=\{10, 12\}\tau_B$),
we observed a similar shift in the (weaker) oscillations and a marked  
dependence on the
initial conditions. This is shown in Fig.~\ref{fig.oscillating_wells_T2.5_3},
at different temperatures (different symbols) and initial conditions (top and
bottom curves, respectively).  The timescales of the relaxation mechanisms
(bond break/formation and translation/rotational diffusion) depend strongly on
temperature. Thus, at sufficiently low temperatures, the rate of change of
$R_\mathrm{W}$ is too fast preventing the particles to relax to the structures 
expected for the corresponding static traps $R_\mathrm{W}$.

\section{Conclusions}\label{sec.conclusions} 

We have studied the dynamics of spherical colloidal particles on a surface in the 
presence of circular potential traps. The attractive interaction between the particles has
three-fold symmetry. However, in the presence of potential traps, we observe a
crossover from the expected local three-fold symmetry to a six-fold one when
the width of the traps is reduced. For intermediate values of the trap width, we find a 
core-shell structure, where the symmetry in the core is six
fold, while that in the shell is three fold. Note that, for a fixed distance
between the centers of the traps, increasing the width of the traps corresponds
to smoothing out the external potential landscape. Thus, we expect similar results
when the strength of the potential is changed, at fixed width.

For traps with oscillating widths, we find that the final structure may deviate significantly 
from the thermodynamic one, when the relaxation timescales are comparable to the 
period of the oscillations.
In this limit, the final structures depend on the rate of change of the trap width, the 
thermostat temperature, and the initial conditions.

\acknowledgments{ This work was partially funded by the Portuguese German
FCT/DAAD project ``Self-organization of colloidal particles on patterned
substrate'' (DAAD project id: 57339919). We also acknowledge financial support
from the Portuguese Foundation for Science and Technology (FCT) under Contracts
nos. PTDC/FIS-MAC/28146/2017, UID/FIS/00618/2013, and SFRH/BPD/114839/2016.  }

\bibliography{/home/cris/Docs/Research/Articles/Bibtex/SoftMatter}

\begin{thebibliography}{49}%
\makeatletter
\providecommand \@ifxundefined [1]{%
 \@ifx{#1\undefined}
}%
\providecommand \@ifnum [1]{%
 \ifnum #1\expandafter \@firstoftwo
 \else \expandafter \@secondoftwo
 \fi
}%
\providecommand \@ifx [1]{%
 \ifx #1\expandafter \@firstoftwo
 \else \expandafter \@secondoftwo
 \fi
}%
\providecommand \natexlab [1]{#1}%
\providecommand \enquote  [1]{``#1''}%
\providecommand \bibnamefont  [1]{#1}%
\providecommand \bibfnamefont [1]{#1}%
\providecommand \citenamefont [1]{#1}%
\providecommand \href@noop [0]{\@secondoftwo}%
\providecommand \href [0]{\begingroup \@sanitize@url \@href}%
\providecommand \@href[1]{\@@startlink{#1}\@@href}%
\providecommand \@@href[1]{\endgroup#1\@@endlink}%
\providecommand \@sanitize@url [0]{\catcode `\\12\catcode `\$12\catcode
  `\&12\catcode `\#12\catcode `\^12\catcode `\_12\catcode `\%12\relax}%
\providecommand \@@startlink[1]{}%
\providecommand \@@endlink[0]{}%
\providecommand \url  [0]{\begingroup\@sanitize@url \@url }%
\providecommand \@url [1]{\endgroup\@href {#1}{\urlprefix }}%
\providecommand \urlprefix  [0]{URL }%
\providecommand \Eprint [0]{\href }%
\providecommand \doibase [0]{http://dx.doi.org/}%
\providecommand \selectlanguage [0]{\@gobble}%
\providecommand \bibinfo  [0]{\@secondoftwo}%
\providecommand \bibfield  [0]{\@secondoftwo}%
\providecommand \translation [1]{[#1]}%
\providecommand \BibitemOpen [0]{}%
\providecommand \bibitemStop [0]{}%
\providecommand \bibitemNoStop [0]{.\EOS\space}%
\providecommand \EOS [0]{\spacefactor3000\relax}%
\providecommand \BibitemShut  [1]{\csname bibitem#1\endcsname}%
\let\auto@bib@innerbib\@empty
\bibitem [{\citenamefont {Sacanna}\ and\ \citenamefont
  {Pine}(2011)}]{Sacanna2011}%
  \BibitemOpen
  \bibfield  {author} {\bibinfo {author} {\bibfnamefont {S.}~\bibnamefont
  {Sacanna}}\ and\ \bibinfo {author} {\bibfnamefont {D.~J.}\ \bibnamefont
  {Pine}},\ }\href@noop {} {\bibfield  {journal} {\bibinfo  {journal} {Curr.
  Op. Coll. Interf. Sci.}\ }\textbf {\bibinfo {volume} {16}},\ \bibinfo {pages}
  {96} (\bibinfo {year} {2011})}\BibitemShut {NoStop}%
\bibitem [{\citenamefont {Chen}\ \emph {et~al.}(2011)\citenamefont {Chen},
  \citenamefont {Bae},\ and\ \citenamefont {Granick}}]{Chen2011}%
  \BibitemOpen
  \bibfield  {author} {\bibinfo {author} {\bibfnamefont {Q.}~\bibnamefont
  {Chen}}, \bibinfo {author} {\bibfnamefont {S.~C.}\ \bibnamefont {Bae}}, \
  and\ \bibinfo {author} {\bibfnamefont {S.}~\bibnamefont {Granick}},\
  }\href@noop {} {\bibfield  {journal} {\bibinfo  {journal} {Nature}\ }\textbf
  {\bibinfo {volume} {469}},\ \bibinfo {pages} {381} (\bibinfo {year}
  {2011})}\BibitemShut {NoStop}%
\bibitem [{\citenamefont {Romano}\ and\ \citenamefont
  {Sciortino}(2011)}]{Romano2011a}%
  \BibitemOpen
  \bibfield  {author} {\bibinfo {author} {\bibfnamefont {F.}~\bibnamefont
  {Romano}}\ and\ \bibinfo {author} {\bibfnamefont {F.}~\bibnamefont
  {Sciortino}},\ }\href@noop {} {\bibfield  {journal} {\bibinfo  {journal}
  {Nat. Mater.}\ }\textbf {\bibinfo {volume} {10}},\ \bibinfo {pages} {171}
  (\bibinfo {year} {2011})}\BibitemShut {NoStop}%
\bibitem [{\citenamefont {Nykypanchuk}\ \emph {et~al.}(2008)\citenamefont
  {Nykypanchuk}, \citenamefont {Maye}, \citenamefont {van~der Lelie},\ and\
  \citenamefont {Gang}}]{Nykypanchuk2008}%
  \BibitemOpen
  \bibfield  {author} {\bibinfo {author} {\bibfnamefont {D.}~\bibnamefont
  {Nykypanchuk}}, \bibinfo {author} {\bibfnamefont {M.~M.}\ \bibnamefont
  {Maye}}, \bibinfo {author} {\bibfnamefont {D.}~\bibnamefont {van~der Lelie}},
  \ and\ \bibinfo {author} {\bibfnamefont {O.}~\bibnamefont {Gang}},\
  }\href@noop {} {\bibfield  {journal} {\bibinfo  {journal} {Nature}\ }\textbf
  {\bibinfo {volume} {451}},\ \bibinfo {pages} {549} (\bibinfo {year}
  {2008})}\BibitemShut {NoStop}%
\bibitem [{\citenamefont {Parak}(2011)}]{Parak2011}%
  \BibitemOpen
  \bibfield  {author} {\bibinfo {author} {\bibfnamefont {W.~J.}\ \bibnamefont
  {Parak}},\ }\href@noop {} {\bibfield  {journal} {\bibinfo  {journal}
  {Science}\ }\textbf {\bibinfo {volume} {334}},\ \bibinfo {pages} {1359}
  (\bibinfo {year} {2011})}\BibitemShut {NoStop}%
\bibitem [{\citenamefont {Yang}\ \emph {et~al.}(2008)\citenamefont {Yang},
  \citenamefont {Kim}, \citenamefont {Lim},\ and\ \citenamefont
  {Yi}}]{Yang2008}%
  \BibitemOpen
  \bibfield  {author} {\bibinfo {author} {\bibfnamefont {S.-M.}\ \bibnamefont
  {Yang}}, \bibinfo {author} {\bibfnamefont {S.-H.}\ \bibnamefont {Kim}},
  \bibinfo {author} {\bibfnamefont {J.-M.}\ \bibnamefont {Lim}}, \ and\
  \bibinfo {author} {\bibfnamefont {G.-R.}\ \bibnamefont {Yi}},\ }\href@noop {}
  {\bibfield  {journal} {\bibinfo  {journal} {J. Mater. Chem.}\ }\textbf
  {\bibinfo {volume} {18}},\ \bibinfo {pages} {2177} (\bibinfo {year}
  {2008})}\BibitemShut {NoStop}%
\bibitem [{\citenamefont {Furst}(2011)}]{Furst2011}%
  \BibitemOpen
  \bibfield  {author} {\bibinfo {author} {\bibfnamefont {E.~M.}\ \bibnamefont
  {Furst}},\ }\href@noop {} {\bibfield  {journal} {\bibinfo  {journal} {Proc.
  Natl. Acad. Sci.}\ }\textbf {\bibinfo {volume} {108}},\ \bibinfo {pages}
  {20853} (\bibinfo {year} {2011})}\BibitemShut {NoStop}%
\bibitem [{\citenamefont {Manoharan}(2015)}]{Manoharan2015}%
  \BibitemOpen
  \bibfield  {author} {\bibinfo {author} {\bibfnamefont {V.~N.}\ \bibnamefont
  {Manoharan}},\ }\href@noop {} {\bibfield  {journal} {\bibinfo  {journal}
  {Science}\ }\textbf {\bibinfo {volume} {349}},\ \bibinfo {pages} {1253751}
  (\bibinfo {year} {2015})}\BibitemShut {NoStop}%
\bibitem [{\citenamefont {Duguet}\ \emph {et~al.}(2011)\citenamefont {Duguet},
  \citenamefont {D{\'{e}}sert}, \citenamefont {Perro},\ and\ \citenamefont
  {Ravaine}}]{Duguet2011}%
  \BibitemOpen
  \bibfield  {author} {\bibinfo {author} {\bibfnamefont {E.}~\bibnamefont
  {Duguet}}, \bibinfo {author} {\bibfnamefont {A.}~\bibnamefont
  {D{\'{e}}sert}}, \bibinfo {author} {\bibfnamefont {A.}~\bibnamefont {Perro}},
  \ and\ \bibinfo {author} {\bibfnamefont {S.}~\bibnamefont {Ravaine}},\
  }\href@noop {} {\bibfield  {journal} {\bibinfo  {journal} {Chem. Soc. Rev.}\
  }\textbf {\bibinfo {volume} {40}},\ \bibinfo {pages} {941} (\bibinfo {year}
  {2011})}\BibitemShut {NoStop}%
\bibitem [{\citenamefont {Dias}\ \emph {et~al.}(2017)\citenamefont {Dias},
  \citenamefont {Ara{\'{u}}jo},\ and\ \citenamefont {{Telo da
  Gama}}}]{Dias2017}%
  \BibitemOpen
  \bibfield  {author} {\bibinfo {author} {\bibfnamefont {C.~S.}\ \bibnamefont
  {Dias}}, \bibinfo {author} {\bibfnamefont {N.~A.~M.}\ \bibnamefont
  {Ara{\'{u}}jo}}, \ and\ \bibinfo {author} {\bibfnamefont {M.~M.}\
  \bibnamefont {{Telo da Gama}}},\ }\href@noop {} {\bibfield  {journal}
  {\bibinfo  {journal} {Adv. Col. Interf. Sci.}\ }\textbf {\bibinfo {volume}
  {247}},\ \bibinfo {pages} {258} (\bibinfo {year} {2017})}\BibitemShut
  {NoStop}%
\bibitem [{\citenamefont {Ramsteiner}\ \emph {et~al.}(2009)\citenamefont
  {Ramsteiner}, \citenamefont {Jensen}, \citenamefont {Weitz},\ and\
  \citenamefont {Spaepen}}]{Ramsteiner2009}%
  \BibitemOpen
  \bibfield  {author} {\bibinfo {author} {\bibfnamefont {I.~B.}\ \bibnamefont
  {Ramsteiner}}, \bibinfo {author} {\bibfnamefont {K.~E.}\ \bibnamefont
  {Jensen}}, \bibinfo {author} {\bibfnamefont {D.~A.}\ \bibnamefont {Weitz}}, \
  and\ \bibinfo {author} {\bibfnamefont {F.}~\bibnamefont {Spaepen}},\
  }\href@noop {} {\bibfield  {journal} {\bibinfo  {journal} {Phys. Rev. E}\
  }\textbf {\bibinfo {volume} {79}},\ \bibinfo {pages} {011403} (\bibinfo
  {year} {2009})}\BibitemShut {NoStop}%
\bibitem [{\citenamefont {Wang}\ and\ \citenamefont
  {M{\"{o}}hwald}(2004)}]{Wang2004a}%
  \BibitemOpen
  \bibfield  {author} {\bibinfo {author} {\bibfnamefont {D.}~\bibnamefont
  {Wang}}\ and\ \bibinfo {author} {\bibfnamefont {H.}~\bibnamefont
  {M{\"{o}}hwald}},\ }\href@noop {} {\bibfield  {journal} {\bibinfo  {journal}
  {J. Mater. Chem.}\ }\textbf {\bibinfo {volume} {14}},\ \bibinfo {pages} {459}
  (\bibinfo {year} {2004})}\BibitemShut {NoStop}%
\bibitem [{\citenamefont {Tierno}\ and\ \citenamefont
  {Fischer}(2014)}]{Tierno2014}%
  \BibitemOpen
  \bibfield  {author} {\bibinfo {author} {\bibfnamefont {P.}~\bibnamefont
  {Tierno}}\ and\ \bibinfo {author} {\bibfnamefont {T.~M.}\ \bibnamefont
  {Fischer}},\ }\href@noop {} {\bibfield  {journal} {\bibinfo  {journal} {Phys.
  Rev. Lett.}\ }\textbf {\bibinfo {volume} {112}},\ \bibinfo {pages} {048302}
  (\bibinfo {year} {2014})}\BibitemShut {NoStop}%
\bibitem [{\citenamefont {Loehr}\ \emph {et~al.}(2018)\citenamefont {Loehr},
  \citenamefont {de~las Heras}, \citenamefont {Jarosz}, \citenamefont
  {Urbaniak}, \citenamefont {Stobiecki}, \citenamefont {Tomita}, \citenamefont
  {Huhnstock}, \citenamefont {Koch}, \citenamefont {Ehresmann}, \citenamefont
  {Holzinger},\ and\ \citenamefont {Fischer}}]{Loehr2017}%
  \BibitemOpen
  \bibfield  {author} {\bibinfo {author} {\bibfnamefont {J.}~\bibnamefont
  {Loehr}}, \bibinfo {author} {\bibfnamefont {D.}~\bibnamefont {de~las Heras}},
  \bibinfo {author} {\bibfnamefont {A.}~\bibnamefont {Jarosz}}, \bibinfo
  {author} {\bibfnamefont {M.}~\bibnamefont {Urbaniak}}, \bibinfo {author}
  {\bibfnamefont {F.}~\bibnamefont {Stobiecki}}, \bibinfo {author}
  {\bibfnamefont {A.}~\bibnamefont {Tomita}}, \bibinfo {author} {\bibfnamefont
  {R.}~\bibnamefont {Huhnstock}}, \bibinfo {author} {\bibfnamefont
  {I.}~\bibnamefont {Koch}}, \bibinfo {author} {\bibfnamefont {A.}~\bibnamefont
  {Ehresmann}}, \bibinfo {author} {\bibfnamefont {D.}~\bibnamefont
  {Holzinger}}, \ and\ \bibinfo {author} {\bibfnamefont {T.~M.}\ \bibnamefont
  {Fischer}},\ }\href@noop {} {\bibfield  {journal} {\bibinfo  {journal} {Comm.
  Phys.}\ }\textbf {\bibinfo {volume} {1}},\ \bibinfo {pages} {4} (\bibinfo
  {year} {2018})}\BibitemShut {NoStop}%
\bibitem [{\citenamefont {Aizenberg}\ \emph {et~al.}(1999)\citenamefont
  {Aizenberg}, \citenamefont {Black},\ and\ \citenamefont
  {Whitesides}}]{Aizenberg1999}%
  \BibitemOpen
  \bibfield  {author} {\bibinfo {author} {\bibfnamefont {J.}~\bibnamefont
  {Aizenberg}}, \bibinfo {author} {\bibfnamefont {A.~J.}\ \bibnamefont
  {Black}}, \ and\ \bibinfo {author} {\bibfnamefont {G.~M.}\ \bibnamefont
  {Whitesides}},\ }\href {\doibase 10.1038/19047} {\bibfield  {journal}
  {\bibinfo  {journal} {Nature}\ }\textbf {\bibinfo {volume} {398}},\ \bibinfo
  {pages} {495} (\bibinfo {year} {1999})}\BibitemShut {NoStop}%
\bibitem [{\citenamefont {Chen}\ \emph {et~al.}(2000)\citenamefont {Chen},
  \citenamefont {Jiang}, \citenamefont {Kimerling},\ and\ \citenamefont
  {Hammond}}]{Chen2000}%
  \BibitemOpen
  \bibfield  {author} {\bibinfo {author} {\bibfnamefont {K.~M.}\ \bibnamefont
  {Chen}}, \bibinfo {author} {\bibfnamefont {X.}~\bibnamefont {Jiang}},
  \bibinfo {author} {\bibfnamefont {L.~C.}\ \bibnamefont {Kimerling}}, \ and\
  \bibinfo {author} {\bibfnamefont {P.~T.}\ \bibnamefont {Hammond}},\ }\href
  {\doibase 10.1021/la000277c} {\bibfield  {journal} {\bibinfo  {journal}
  {Langmuir}\ }\textbf {\bibinfo {volume} {16}},\ \bibinfo {pages} {7825}
  (\bibinfo {year} {2000})}\BibitemShut {NoStop}%
\bibitem [{\citenamefont {Guo}\ \emph {et~al.}(2001)\citenamefont {Guo},
  \citenamefont {Arnoux},\ and\ \citenamefont {Palmer}}]{Guo2001}%
  \BibitemOpen
  \bibfield  {author} {\bibinfo {author} {\bibfnamefont {Q.}~\bibnamefont
  {Guo}}, \bibinfo {author} {\bibfnamefont {C.}~\bibnamefont {Arnoux}}, \ and\
  \bibinfo {author} {\bibfnamefont {R.~E.}\ \bibnamefont {Palmer}},\
  }\href@noop {} {\bibfield  {journal} {\bibinfo  {journal} {Langmuir}\
  }\textbf {\bibinfo {volume} {17}},\ \bibinfo {pages} {7150} (\bibinfo {year}
  {2001})}\BibitemShut {NoStop}%
\bibitem [{\citenamefont {Joshi}\ \emph {et~al.}(2016)\citenamefont {Joshi},
  \citenamefont {Bargteil}, \citenamefont {Caciagli}, \citenamefont
  {Burelbach}, \citenamefont {Xing}, \citenamefont {Nunes}, \citenamefont
  {Pinto}, \citenamefont {Ara{\'{u}}jo}, \citenamefont {Bruijc},\ and\
  \citenamefont {Eiser}}]{Joshi2016}%
  \BibitemOpen
  \bibfield  {author} {\bibinfo {author} {\bibfnamefont {D.}~\bibnamefont
  {Joshi}}, \bibinfo {author} {\bibfnamefont {D.}~\bibnamefont {Bargteil}},
  \bibinfo {author} {\bibfnamefont {A.}~\bibnamefont {Caciagli}}, \bibinfo
  {author} {\bibfnamefont {J.}~\bibnamefont {Burelbach}}, \bibinfo {author}
  {\bibfnamefont {Z.}~\bibnamefont {Xing}}, \bibinfo {author} {\bibfnamefont
  {A.~S.}\ \bibnamefont {Nunes}}, \bibinfo {author} {\bibfnamefont {D.~E.~P.}\
  \bibnamefont {Pinto}}, \bibinfo {author} {\bibfnamefont {N.~A.~M.}\
  \bibnamefont {Ara{\'{u}}jo}}, \bibinfo {author} {\bibfnamefont
  {J.}~\bibnamefont {Bruijc}}, \ and\ \bibinfo {author} {\bibfnamefont
  {E.}~\bibnamefont {Eiser}},\ }\href@noop {} {\bibfield  {journal} {\bibinfo
  {journal} {Sci. Adv.}\ }\textbf {\bibinfo {volume} {2}},\ \bibinfo {pages}
  {e1600881} (\bibinfo {year} {2016})}\BibitemShut {NoStop}%
\bibitem [{\citenamefont {Heni}\ and\ \citenamefont
  {L{\"{o}}wen}(2000)}]{Heni2000}%
  \BibitemOpen
  \bibfield  {author} {\bibinfo {author} {\bibfnamefont {M.}~\bibnamefont
  {Heni}}\ and\ \bibinfo {author} {\bibfnamefont {H.}~\bibnamefont
  {L{\"{o}}wen}},\ }\href@noop {} {\bibfield  {journal} {\bibinfo  {journal}
  {Phys. Rev. Lett.}\ }\textbf {\bibinfo {volume} {85}},\ \bibinfo {pages}
  {3668} (\bibinfo {year} {2000})}\BibitemShut {NoStop}%
\bibitem [{\citenamefont {Harreis}\ \emph {et~al.}(2002)\citenamefont
  {Harreis}, \citenamefont {Schmidt},\ and\ \citenamefont
  {L{\"{o}}wen}}]{Harreis2002}%
  \BibitemOpen
  \bibfield  {author} {\bibinfo {author} {\bibfnamefont {H.~M.}\ \bibnamefont
  {Harreis}}, \bibinfo {author} {\bibfnamefont {M.}~\bibnamefont {Schmidt}}, \
  and\ \bibinfo {author} {\bibfnamefont {H.}~\bibnamefont {L{\"{o}}wen}},\
  }\href@noop {} {\bibfield  {journal} {\bibinfo  {journal} {Phys. Rev. E}\
  }\textbf {\bibinfo {volume} {65}},\ \bibinfo {pages} {041602} (\bibinfo
  {year} {2002})}\BibitemShut {NoStop}%
\bibitem [{\citenamefont {Bauer}\ and\ \citenamefont
  {Dietrich}(1999)}]{Bauer1999a}%
  \BibitemOpen
  \bibfield  {author} {\bibinfo {author} {\bibfnamefont {C.}~\bibnamefont
  {Bauer}}\ and\ \bibinfo {author} {\bibfnamefont {S.}~\bibnamefont
  {Dietrich}},\ }\href@noop {} {\bibfield  {journal} {\bibinfo  {journal}
  {Phys. Rev. E}\ }\textbf {\bibinfo {volume} {60}},\ \bibinfo {pages} {6919}
  (\bibinfo {year} {1999})}\BibitemShut {NoStop}%
\bibitem [{\citenamefont {Cadilhe}\ \emph {et~al.}(2007)\citenamefont
  {Cadilhe}, \citenamefont {Ara{\'{u}}jo},\ and\ \citenamefont
  {Privman}}]{Cadilhe2007}%
  \BibitemOpen
  \bibfield  {author} {\bibinfo {author} {\bibfnamefont {A.}~\bibnamefont
  {Cadilhe}}, \bibinfo {author} {\bibfnamefont {N.~A.~M.}\ \bibnamefont
  {Ara{\'{u}}jo}}, \ and\ \bibinfo {author} {\bibfnamefont {V.}~\bibnamefont
  {Privman}},\ }\href@noop {} {\bibfield  {journal} {\bibinfo  {journal} {J.
  Phys.: Condens. Matter}\ }\textbf {\bibinfo {volume} {19}},\ \bibinfo {pages}
  {065124} (\bibinfo {year} {2007})}\BibitemShut {NoStop}%
\bibitem [{\citenamefont {Sacanna}\ \emph {et~al.}(2013)\citenamefont
  {Sacanna}, \citenamefont {Pine},\ and\ \citenamefont {Yi}}]{Sacanna2013a}%
  \BibitemOpen
  \bibfield  {author} {\bibinfo {author} {\bibfnamefont {S.}~\bibnamefont
  {Sacanna}}, \bibinfo {author} {\bibfnamefont {D.~J.}\ \bibnamefont {Pine}}, \
  and\ \bibinfo {author} {\bibfnamefont {G.-R.}\ \bibnamefont {Yi}},\
  }\href@noop {} {\bibfield  {journal} {\bibinfo  {journal} {Soft Matt.}\
  }\textbf {\bibinfo {volume} {9}},\ \bibinfo {pages} {8096} (\bibinfo {year}
  {2013})}\BibitemShut {NoStop}%
\bibitem [{\citenamefont {Wolters}\ \emph {et~al.}(2015)\citenamefont
  {Wolters}, \citenamefont {Avvisati}, \citenamefont {Hagemans}, \citenamefont
  {Vissers}, \citenamefont {Kraft}, \citenamefont {Dijkstra},\ and\
  \citenamefont {Kegel}}]{Wolters2015}%
  \BibitemOpen
  \bibfield  {author} {\bibinfo {author} {\bibfnamefont {J.~R.}\ \bibnamefont
  {Wolters}}, \bibinfo {author} {\bibfnamefont {G.}~\bibnamefont {Avvisati}},
  \bibinfo {author} {\bibfnamefont {F.}~\bibnamefont {Hagemans}}, \bibinfo
  {author} {\bibfnamefont {T.}~\bibnamefont {Vissers}}, \bibinfo {author}
  {\bibfnamefont {D.~J.}\ \bibnamefont {Kraft}}, \bibinfo {author}
  {\bibfnamefont {M.}~\bibnamefont {Dijkstra}}, \ and\ \bibinfo {author}
  {\bibfnamefont {W.~K.}\ \bibnamefont {Kegel}},\ }\href@noop {} {\bibfield
  {journal} {\bibinfo  {journal} {Soft Matt.}\ }\textbf {\bibinfo {volume}
  {11}},\ \bibinfo {pages} {1067} (\bibinfo {year} {2015})}\BibitemShut
  {NoStop}%
\bibitem [{\citenamefont {Glotzer}(2012)}]{Glotzer2012}%
  \BibitemOpen
  \bibfield  {author} {\bibinfo {author} {\bibfnamefont {S.~C.}\ \bibnamefont
  {Glotzer}},\ }\href@noop {} {\bibfield  {journal} {\bibinfo  {journal}
  {Nature}\ }\textbf {\bibinfo {volume} {481}},\ \bibinfo {pages} {450}
  (\bibinfo {year} {2012})}\BibitemShut {NoStop}%
\bibitem [{\citenamefont {Yunker}\ \emph {et~al.}(2013)\citenamefont {Yunker},
  \citenamefont {Lohr}, \citenamefont {Still}, \citenamefont {Borodin},
  \citenamefont {Durian},\ and\ \citenamefont {Yodh}}]{Yunker2013}%
  \BibitemOpen
  \bibfield  {author} {\bibinfo {author} {\bibfnamefont {P.~J.}\ \bibnamefont
  {Yunker}}, \bibinfo {author} {\bibfnamefont {M.~A.}\ \bibnamefont {Lohr}},
  \bibinfo {author} {\bibfnamefont {T.}~\bibnamefont {Still}}, \bibinfo
  {author} {\bibfnamefont {A.}~\bibnamefont {Borodin}}, \bibinfo {author}
  {\bibfnamefont {D.~J.}\ \bibnamefont {Durian}}, \ and\ \bibinfo {author}
  {\bibfnamefont {A.~G.}\ \bibnamefont {Yodh}},\ }\href@noop {} {\bibfield
  {journal} {\bibinfo  {journal} {Phys. Rev. Lett.}\ }\textbf {\bibinfo
  {volume} {110}},\ \bibinfo {pages} {035501} (\bibinfo {year}
  {2013})}\BibitemShut {NoStop}%
\bibitem [{\citenamefont {Dias}\ \emph
  {et~al.}(2018{\natexlab{a}})\citenamefont {Dias}, \citenamefont {Yunker},
  \citenamefont {Yodh}, \citenamefont {Ara{\'{u}}jo},\ and\ \citenamefont
  {{Telo da Gama}}}]{Dias2018a}%
  \BibitemOpen
  \bibfield  {author} {\bibinfo {author} {\bibfnamefont {C.~S.}\ \bibnamefont
  {Dias}}, \bibinfo {author} {\bibfnamefont {P.~J.}\ \bibnamefont {Yunker}},
  \bibinfo {author} {\bibfnamefont {A.~G.}\ \bibnamefont {Yodh}}, \bibinfo
  {author} {\bibfnamefont {N.~A.~M.}\ \bibnamefont {Ara{\'{u}}jo}}, \ and\
  \bibinfo {author} {\bibfnamefont {M.~M.}\ \bibnamefont {{Telo da Gama}}},\
  }\href@noop {} {\bibfield  {journal} {\bibinfo  {journal} {Soft Matt.}\
  }\textbf {\bibinfo {volume} {14}},\ \bibinfo {pages} {1903} (\bibinfo {year}
  {2018}{\natexlab{a}})}\BibitemShut {NoStop}%
\bibitem [{\citenamefont {Ilg}\ and\ \citenamefont {{Del
  Gado}}(2011)}]{Ilg2011}%
  \BibitemOpen
  \bibfield  {author} {\bibinfo {author} {\bibfnamefont {P.}~\bibnamefont
  {Ilg}}\ and\ \bibinfo {author} {\bibfnamefont {E.}~\bibnamefont {{Del
  Gado}}},\ }\href@noop {} {\bibfield  {journal} {\bibinfo  {journal} {Soft
  Matt.}\ }\textbf {\bibinfo {volume} {7}},\ \bibinfo {pages} {163} (\bibinfo
  {year} {2011})}\BibitemShut {NoStop}%
\bibitem [{\citenamefont {Nych}\ \emph {et~al.}(2013)\citenamefont {Nych},
  \citenamefont {Ognysta}, \citenamefont {Skarabot}, \citenamefont {Ravnik},
  \citenamefont {Zumer},\ and\ \citenamefont {Mu{\v{s}}evi{\v{c}}}}]{Nych2013}%
  \BibitemOpen
  \bibfield  {author} {\bibinfo {author} {\bibfnamefont {A.}~\bibnamefont
  {Nych}}, \bibinfo {author} {\bibfnamefont {U.}~\bibnamefont {Ognysta}},
  \bibinfo {author} {\bibfnamefont {M.}~\bibnamefont {Skarabot}}, \bibinfo
  {author} {\bibfnamefont {M.}~\bibnamefont {Ravnik}}, \bibinfo {author}
  {\bibfnamefont {S.}~\bibnamefont {Zumer}}, \ and\ \bibinfo {author}
  {\bibfnamefont {I.}~\bibnamefont {Mu{\v{s}}evi{\v{c}}}},\ }\href@noop {}
  {\bibfield  {journal} {\bibinfo  {journal} {Nat. Commun.}\ }\textbf {\bibinfo
  {volume} {4}},\ \bibinfo {pages} {1489} (\bibinfo {year} {2013})}\BibitemShut
  {NoStop}%
\bibitem [{\citenamefont {Klapp}(2016)}]{Klapp2016}%
  \BibitemOpen
  \bibfield  {author} {\bibinfo {author} {\bibfnamefont {S.~H.~L.}\
  \bibnamefont {Klapp}},\ }\href@noop {} {\bibfield  {journal} {\bibinfo
  {journal} {Curr. Op. Coll. Interf. Sci.}\ }\textbf {\bibinfo {volume} {21}},\
  \bibinfo {pages} {76} (\bibinfo {year} {2016})}\BibitemShut {NoStop}%
\bibitem [{\citenamefont {Wang}\ \emph {et~al.}(2012)\citenamefont {Wang},
  \citenamefont {Wang}, \citenamefont {Breed}, \citenamefont {Manoharan},
  \citenamefont {Feng}, \citenamefont {Hollingsworth}, \citenamefont {Weck},\
  and\ \citenamefont {Pine}}]{Wang2012}%
  \BibitemOpen
  \bibfield  {author} {\bibinfo {author} {\bibfnamefont {Y.}~\bibnamefont
  {Wang}}, \bibinfo {author} {\bibfnamefont {Y.}~\bibnamefont {Wang}}, \bibinfo
  {author} {\bibfnamefont {D.~R.}\ \bibnamefont {Breed}}, \bibinfo {author}
  {\bibfnamefont {V.~N.}\ \bibnamefont {Manoharan}}, \bibinfo {author}
  {\bibfnamefont {L.}~\bibnamefont {Feng}}, \bibinfo {author} {\bibfnamefont
  {A.~D.}\ \bibnamefont {Hollingsworth}}, \bibinfo {author} {\bibfnamefont
  {M.}~\bibnamefont {Weck}}, \ and\ \bibinfo {author} {\bibfnamefont {D.~J.}\
  \bibnamefont {Pine}},\ }\href@noop {} {\bibfield  {journal} {\bibinfo
  {journal} {Nature}\ }\textbf {\bibinfo {volume} {491}},\ \bibinfo {pages}
  {51} (\bibinfo {year} {2012})}\BibitemShut {NoStop}%
\bibitem [{\citenamefont {Smallenburg}\ \emph {et~al.}(2013)\citenamefont
  {Smallenburg}, \citenamefont {Leibler},\ and\ \citenamefont
  {Sciortino}}]{Smallenburg2013a}%
  \BibitemOpen
  \bibfield  {author} {\bibinfo {author} {\bibfnamefont {F.}~\bibnamefont
  {Smallenburg}}, \bibinfo {author} {\bibfnamefont {L.}~\bibnamefont
  {Leibler}}, \ and\ \bibinfo {author} {\bibfnamefont {F.}~\bibnamefont
  {Sciortino}},\ }\href@noop {} {\bibfield  {journal} {\bibinfo  {journal}
  {Phys. Rev. Lett.}\ }\textbf {\bibinfo {volume} {111}},\ \bibinfo {pages}
  {188002} (\bibinfo {year} {2013})}\BibitemShut {NoStop}%
\bibitem [{\citenamefont {Bianchi}\ \emph {et~al.}(2011)\citenamefont
  {Bianchi}, \citenamefont {Kahl},\ and\ \citenamefont {Likos}}]{Bianchi2011a}%
  \BibitemOpen
  \bibfield  {author} {\bibinfo {author} {\bibfnamefont {E.}~\bibnamefont
  {Bianchi}}, \bibinfo {author} {\bibfnamefont {G.}~\bibnamefont {Kahl}}, \
  and\ \bibinfo {author} {\bibfnamefont {C.~N.}\ \bibnamefont {Likos}},\
  }\href@noop {} {\bibfield  {journal} {\bibinfo  {journal} {Soft Matt.}\
  }\textbf {\bibinfo {volume} {7}},\ \bibinfo {pages} {8313} (\bibinfo {year}
  {2011})}\BibitemShut {NoStop}%
\bibitem [{\citenamefont {Dias}\ \emph {et~al.}(2014)\citenamefont {Dias},
  \citenamefont {Ara{\'{u}}jo},\ and\ \citenamefont {{Telo da
  Gama}}}]{Dias2014a}%
  \BibitemOpen
  \bibfield  {author} {\bibinfo {author} {\bibfnamefont {C.~S.}\ \bibnamefont
  {Dias}}, \bibinfo {author} {\bibfnamefont {N.~A.~M.}\ \bibnamefont
  {Ara{\'{u}}jo}}, \ and\ \bibinfo {author} {\bibfnamefont {M.~M.}\
  \bibnamefont {{Telo da Gama}}},\ }\href@noop {} {\bibfield  {journal}
  {\bibinfo  {journal} {EPL}\ }\textbf {\bibinfo {volume} {107}},\ \bibinfo
  {pages} {56002} (\bibinfo {year} {2014})}\BibitemShut {NoStop}%
\bibitem [{\citenamefont {Soko{\l}owski}\ and\ \citenamefont
  {Kalyuzhnyi}(2014)}]{Sokolowski2014}%
  \BibitemOpen
  \bibfield  {author} {\bibinfo {author} {\bibfnamefont {S.}~\bibnamefont
  {Soko{\l}owski}}\ and\ \bibinfo {author} {\bibfnamefont {Y.~V.}\ \bibnamefont
  {Kalyuzhnyi}},\ }\href@noop {} {\bibfield  {journal} {\bibinfo  {journal} {J.
  Phys. Chem. B}\ }\textbf {\bibinfo {volume} {118}},\ \bibinfo {pages} {9076}
  (\bibinfo {year} {2014})}\BibitemShut {NoStop}%
\bibitem [{\citenamefont {S{\l}yk}\ \emph {et~al.}(2016)\citenamefont
  {S{\l}yk}, \citenamefont {R{\.{z}}ysko},\ and\ \citenamefont
  {Bryk}}]{Syk2016}%
  \BibitemOpen
  \bibfield  {author} {\bibinfo {author} {\bibfnamefont {E.}~\bibnamefont
  {S{\l}yk}}, \bibinfo {author} {\bibfnamefont {W.}~\bibnamefont
  {R{\.{z}}ysko}}, \ and\ \bibinfo {author} {\bibfnamefont {P.}~\bibnamefont
  {Bryk}},\ }\href@noop {} {\bibfield  {journal} {\bibinfo  {journal} {Soft
  Matt.}\ }\textbf {\bibinfo {volume} {12}},\ \bibinfo {pages} {9538} (\bibinfo
  {year} {2016})}\BibitemShut {NoStop}%
\bibitem [{\citenamefont {Yi}\ \emph {et~al.}(2013)\citenamefont {Yi},
  \citenamefont {Pine},\ and\ \citenamefont {Sacanna}}]{Yi2013}%
  \BibitemOpen
  \bibfield  {author} {\bibinfo {author} {\bibfnamefont {G.-R.}\ \bibnamefont
  {Yi}}, \bibinfo {author} {\bibfnamefont {D.~J.}\ \bibnamefont {Pine}}, \ and\
  \bibinfo {author} {\bibfnamefont {S.}~\bibnamefont {Sacanna}},\ }\href@noop
  {} {\bibfield  {journal} {\bibinfo  {journal} {J. Phys.: Condens. Matter}\
  }\textbf {\bibinfo {volume} {25}},\ \bibinfo {pages} {193101} (\bibinfo
  {year} {2013})}\BibitemShut {NoStop}%
\bibitem [{\citenamefont {He}\ and\ \citenamefont
  {Kretzschmar}(2012)}]{He2012}%
  \BibitemOpen
  \bibfield  {author} {\bibinfo {author} {\bibfnamefont {Z.}~\bibnamefont
  {He}}\ and\ \bibinfo {author} {\bibfnamefont {I.}~\bibnamefont
  {Kretzschmar}},\ }\href@noop {} {\bibfield  {journal} {\bibinfo  {journal}
  {Langmuir}\ }\textbf {\bibinfo {volume} {28}},\ \bibinfo {pages} {9915}
  (\bibinfo {year} {2012})}\BibitemShut {NoStop}%
\bibitem [{\citenamefont {Kraft}\ \emph {et~al.}(2011)\citenamefont {Kraft},
  \citenamefont {Hilhorst}, \citenamefont {Heinen}, \citenamefont {Hoogenraad},
  \citenamefont {Luigjes},\ and\ \citenamefont {Kegel}}]{Kraft2011}%
  \BibitemOpen
  \bibfield  {author} {\bibinfo {author} {\bibfnamefont {D.~J.}\ \bibnamefont
  {Kraft}}, \bibinfo {author} {\bibfnamefont {J.}~\bibnamefont {Hilhorst}},
  \bibinfo {author} {\bibfnamefont {M.~A.~P.}\ \bibnamefont {Heinen}}, \bibinfo
  {author} {\bibfnamefont {M.~J.}\ \bibnamefont {Hoogenraad}}, \bibinfo
  {author} {\bibfnamefont {B.}~\bibnamefont {Luigjes}}, \ and\ \bibinfo
  {author} {\bibfnamefont {W.~K.}\ \bibnamefont {Kegel}},\ }\href@noop {}
  {\bibfield  {journal} {\bibinfo  {journal} {J. Phys. Chem. B}\ }\textbf
  {\bibinfo {volume} {115}},\ \bibinfo {pages} {7175} (\bibinfo {year}
  {2011})}\BibitemShut {NoStop}%
\bibitem [{\citenamefont {Cates}(2013)}]{Cates2013}%
  \BibitemOpen
  \bibfield  {author} {\bibinfo {author} {\bibfnamefont {M.~E.}\ \bibnamefont
  {Cates}},\ }\href@noop {} {\bibfield  {journal} {\bibinfo  {journal} {Nat.
  Mater.}\ }\textbf {\bibinfo {volume} {12}},\ \bibinfo {pages} {179} (\bibinfo
  {year} {2013})}\BibitemShut {NoStop}%
\bibitem [{\citenamefont {Geigenfeind}\ and\ \citenamefont {de~las
  Heras}(2017)}]{Geigenfeind2016}%
  \BibitemOpen
  \bibfield  {author} {\bibinfo {author} {\bibfnamefont {T.}~\bibnamefont
  {Geigenfeind}}\ and\ \bibinfo {author} {\bibfnamefont {D.}~\bibnamefont
  {de~las Heras}},\ }\href@noop {} {\bibfield  {journal} {\bibinfo  {journal}
  {J. Phys.: Condens. Matter}\ }\textbf {\bibinfo {volume} {29}},\ \bibinfo
  {pages} {064006} (\bibinfo {year} {2017})}\BibitemShut {NoStop}%
\bibitem [{\citenamefont {Treffenst{\"{a}}dt}\ \emph
  {et~al.}(2018)\citenamefont {Treffenst{\"{a}}dt}, \citenamefont
  {Ara{\'{u}}jo},\ and\ \citenamefont {de~las Heras}}]{Treffensta2018}%
  \BibitemOpen
  \bibfield  {author} {\bibinfo {author} {\bibfnamefont {L.~L.}\ \bibnamefont
  {Treffenst{\"{a}}dt}}, \bibinfo {author} {\bibfnamefont {N.~A.~M.}\
  \bibnamefont {Ara{\'{u}}jo}}, \ and\ \bibinfo {author} {\bibfnamefont
  {D.}~\bibnamefont {de~las Heras}},\ }\href@noop {} {\bibfield  {journal}
  {\bibinfo  {journal} {Soft Matter}\ }\textbf {\bibinfo {volume} {14}},\
  \bibinfo {pages} {3572} (\bibinfo {year} {2018})}\BibitemShut {NoStop}%
\bibitem [{\citenamefont {Dias}\ \emph {et~al.}(2016)\citenamefont {Dias},
  \citenamefont {Braga}, \citenamefont {Ara{\'{u}}jo},\ and\ \citenamefont
  {{Telo da Gama}}}]{Dias2016}%
  \BibitemOpen
  \bibfield  {author} {\bibinfo {author} {\bibfnamefont {C.~S.}\ \bibnamefont
  {Dias}}, \bibinfo {author} {\bibfnamefont {C.}~\bibnamefont {Braga}},
  \bibinfo {author} {\bibfnamefont {N.~A.~M.}\ \bibnamefont {Ara{\'{u}}jo}}, \
  and\ \bibinfo {author} {\bibfnamefont {M.~M.}\ \bibnamefont {{Telo da
  Gama}}},\ }\href@noop {} {\bibfield  {journal} {\bibinfo  {journal} {Soft
  Matt.}\ }\textbf {\bibinfo {volume} {12}},\ \bibinfo {pages} {1550} (\bibinfo
  {year} {2016})}\BibitemShut {NoStop}%
\bibitem [{\citenamefont {Plimpton}(1995)}]{Plimpton1995}%
  \BibitemOpen
  \bibfield  {author} {\bibinfo {author} {\bibfnamefont {S.}~\bibnamefont
  {Plimpton}},\ }\href@noop {} {\bibfield  {journal} {\bibinfo  {journal} {J.
  Comp. Phys.}\ }\textbf {\bibinfo {volume} {117}},\ \bibinfo {pages} {1}
  (\bibinfo {year} {1995})}\BibitemShut {NoStop}%
\bibitem [{\citenamefont {Dunweg}\ and\ \citenamefont
  {Wolfgang}(1991)}]{Dunweg1991}%
  \BibitemOpen
  \bibfield  {author} {\bibinfo {author} {\bibfnamefont {B.}~\bibnamefont
  {Dunweg}}\ and\ \bibinfo {author} {\bibfnamefont {P.}~\bibnamefont
  {Wolfgang}},\ }\href@noop {} {\bibfield  {journal} {\bibinfo  {journal} {Int.
  J. Mod. Phys. C}\ }\textbf {\bibinfo {volume} {2}},\ \bibinfo {pages} {817}
  (\bibinfo {year} {1991})}\BibitemShut {NoStop}%
\bibitem [{\citenamefont {Dias}\ \emph
  {et~al.}(2018{\natexlab{b}})\citenamefont {Dias}, \citenamefont
  {Ara{\'{u}}jo},\ and\ \citenamefont {{Telo da Gama}}}]{Dias2018}%
  \BibitemOpen
  \bibfield  {author} {\bibinfo {author} {\bibfnamefont {C.~S.}\ \bibnamefont
  {Dias}}, \bibinfo {author} {\bibfnamefont {N.~A.~M.}\ \bibnamefont
  {Ara{\'{u}}jo}}, \ and\ \bibinfo {author} {\bibfnamefont {M.~M.}\
  \bibnamefont {{Telo da Gama}}},\ }\href@noop {} {\bibfield  {journal}
  {\bibinfo  {journal} {J. Phys.: Condens. Matter}\ }\textbf {\bibinfo {volume}
  {30}},\ \bibinfo {pages} {014001} (\bibinfo {year}
  {2018}{\natexlab{b}})}\BibitemShut {NoStop}%
\bibitem [{\citenamefont {Nunes}\ \emph {et~al.}(2018)\citenamefont {Nunes},
  \citenamefont {Gupta}, \citenamefont {Ara{\'{u}}jo},\ and\ \citenamefont
  {{Telo da Gama}}}]{Nunes2018}%
  \BibitemOpen
  \bibfield  {author} {\bibinfo {author} {\bibfnamefont {A.~S.}\ \bibnamefont
  {Nunes}}, \bibinfo {author} {\bibfnamefont {A.}~\bibnamefont {Gupta}},
  \bibinfo {author} {\bibfnamefont {N.~A.~M.}\ \bibnamefont {Ara{\'{u}}jo}}, \
  and\ \bibinfo {author} {\bibfnamefont {M.~M.}\ \bibnamefont {{Telo da
  Gama}}},\ }\href {\doibase 10.1080/00268976.2018.1464672} {\bibfield
  {journal} {\bibinfo  {journal} {Mol. Phys.}\ } (\bibinfo {year} {2018}),\
  10.1080/00268976.2018.1464672}\BibitemShut {NoStop}%
\bibitem [{\citenamefont {Dias}\ \emph
  {et~al.}(2018{\natexlab{c}})\citenamefont {Dias}, \citenamefont {Tavares},
  \citenamefont {Ara{\'{u}}jo},\ and\ \citenamefont {{Telo Da
  Gama}}}]{Dias2018b}%
  \BibitemOpen
  \bibfield  {author} {\bibinfo {author} {\bibfnamefont {C.~S.}\ \bibnamefont
  {Dias}}, \bibinfo {author} {\bibfnamefont {J.~M.}\ \bibnamefont {Tavares}},
  \bibinfo {author} {\bibfnamefont {N.~A.~M.}\ \bibnamefont {Ara{\'{u}}jo}}, \
  and\ \bibinfo {author} {\bibfnamefont {M.~M.}\ \bibnamefont {{Telo Da
  Gama}}},\ }\href@noop {} {\bibfield  {journal} {\bibinfo  {journal} {Soft
  Matter}\ }\textbf {\bibinfo {volume} {14}},\ \bibinfo {pages} {2744}
  (\bibinfo {year} {2018}{\natexlab{c}})}\BibitemShut {NoStop}%
\bibitem [{\citenamefont {Tavares}\ \emph {et~al.}(2018)\citenamefont
  {Tavares}, \citenamefont {Dias}, \citenamefont {Ara{\'{u}}jo},\ and\
  \citenamefont {{Telo da Gama}}}]{Tavares2018}%
  \BibitemOpen
  \bibfield  {author} {\bibinfo {author} {\bibfnamefont {J.~M.}\ \bibnamefont
  {Tavares}}, \bibinfo {author} {\bibfnamefont {C.~S.}\ \bibnamefont {Dias}},
  \bibinfo {author} {\bibfnamefont {N.~A.~M.}\ \bibnamefont {Ara{\'{u}}jo}}, \
  and\ \bibinfo {author} {\bibfnamefont {M.~M.}\ \bibnamefont {{Telo da
  Gama}}},\ }\href@noop {} {\bibfield  {journal} {\bibinfo  {journal} {J. Phys.
  Chem. B}\ }\textbf {\bibinfo {volume} {122}},\ \bibinfo {pages} {3514}
  (\bibinfo {year} {2018})}\BibitemShut {NoStop}%
\end{thebibliography}%
\end{document}